\documentclass[12pt,a4paper]{article}
\usepackage{times}

\usepackage[textheight=24.7cm,textwidth=18.3cm]{geometry}

\usepackage{graphicx}%
\usepackage{multirow}%
\usepackage{amsmath,amssymb,amsfonts}%
\usepackage{amsthm}%
\usepackage{mathrsfs}%
\usepackage{color}
\usepackage{xurl}
\usepackage{ragged2e}
\usepackage{microtype}
\usepackage[backend=biber,style=numeric, sorting = none, sortcites=true]{biblatex}
\usepackage{ulem}
\newcommand*\patchAmsMathEnvironmentForLineno[1]{
  \expandafter\let\csname old#1\expandafter\endcsname\csname #1\endcsname
  \expandafter\let\csname oldend#1\expandafter\endcsname\csname end#1\endcsname
  \renewenvironment{#1}
     {\linenomath\csname old#1\endcsname}
     {\csname oldend#1\endcsname\endlinenomath}}
\newcommand*\patchBothAmsMathEnvironmentsForLineno[1]{
  \patchAmsMathEnvironmentForLineno{#1}
  \patchAmsMathEnvironmentForLineno{#1*}}
\AtBeginDocument{
\patchBothAmsMathEnvironmentsForLineno{equation}
\patchBothAmsMathEnvironmentsForLineno{align}
\patchBothAmsMathEnvironmentsForLineno{flalign}
\patchBothAmsMathEnvironmentsForLineno{alignat}
\patchBothAmsMathEnvironmentsForLineno{gather}
\patchBothAmsMathEnvironmentsForLineno{multline}
}
\title{Hybridization of pulse and continuous-wave \\
based optical quantum computation}
\author{Tatsuki Sonoyama,$^{1,4,5}$ Tomoki Sano, $^{1}$ Takumi Suzuki,$^{1}$ Kazuma Takahashi,$^{1}$\\
Takefumi Nomura,$^{1}$ Akito Kawasaki,$^{1,4,5}$ Takahiro Kashiwazaki,$^{2}$ Asuka Inoue,$^{2}$ \\Takeshi Umeki,$^{2}$ Masahiro Yabuno,$^{3}$ Shigehito Miki,$^{3}$ Hirotaka Terai,$^{3}$\\
Kan Takase,$^{1,4,5}$ Warit Asavanant,$^{1,4,5}$ Mamoru Endo,$^{1,4}$ and Akira Furusawa$^{1,4,5\ast}$\\
\normalsize{$^{1}$Department of Applied Physics, School of Engineering, The University of Tokyo,}\\
\normalsize{7-3-1 Hongo, Bunkyo-ku, Tokyo 113-8656, Japan}\\
\normalsize{$^{2}$NTT Device Technology Labs, NTT Inc.,}\\
\normalsize{3-1 Morinosato Wakamiya, Atsugi, Kanagawa 243-0198, Japan}\\
\normalsize{$^{3}$Advanced ICT Research Institute, National Institute of Information and Communications Technology,}\\
\normalsize{588-2 Iwaoka, Nishi-ku, Kobe, Hyogo 651-2492, Japan}\\
\normalsize{$^{4}$Optical Quantum Computing Research Team, RIKEN Center for Quantum Computing,}\\
\normalsize{2-1 Hirosawa, Wako, Saitama 351-0198, Japan}\\
\normalsize{$^{5}$OptQC Corp,}\\
\normalsize{1-21-7 Nishi-Ikebukuro, Toshima-ku, Tokyo 171-0021, Japan}\\
\normalsize{To whom correspondence should be addressed; \normalsize$^{\ast}$E-mail: akiraf@ap.t.u-tokyo.ac.jp}
}

\date{\today}
\begin{document}
\baselineskip24pt
\maketitle

{\bf We propose a pulse and continuous wave (CW) hybrid architecture of continuous-variable measurement\-based optical quantum computation utilizing the strengths of both pulsed and CW light.
In this architecture, input and ancillary non-Gaussian quantum states necessary for fault-tolerance and universality are generated with pulsed light, whereas quantum processors including continuous-variable cluster states and homodyne measurement systems are operated with CW light.
This architecture is expected to enable both generation of quantum states with shorter optical wavepackets for ultrafast computation and low-loss manipulation and measurement of these states.
In this study, as a proof-of-principle, an ultrafast homodyne measurement using a CW local oscillator was performed on single-photon states generated with pulsed light.
The measured single-photon state’s temporal width was around 70 ps and the value of the Wigner function at the origin was $W(0,0) = -0.153\pm0.003$, which is highly non-classical.
This will be a core technology for high-speed optical quantum information processing.}

\section*{Introduction}
\begin{figure}[ht]
\centering
 \includegraphics[width=\textwidth]{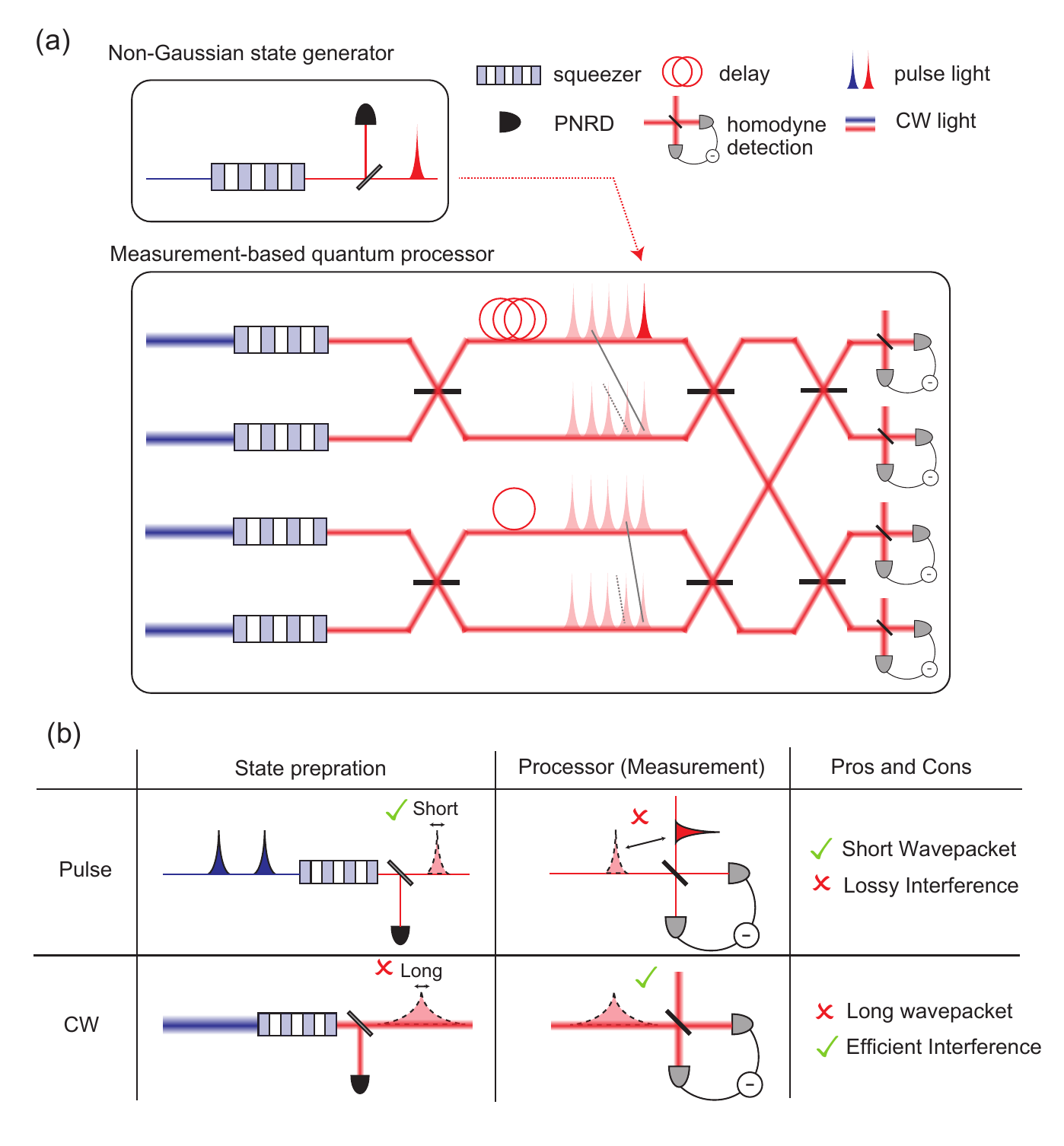}
 \caption{(a) A new optical quantum computing architecture using the pulse-CW hybrid technology experimentally demonstrated in this study. Non-Gaussian states used as initial states or auxiliary states are generated using pulsed light, while the cluster state \cite{RIKENfullstack}, which serves as the computational platform for measurement-based quantum computing, and the homodyne measurement system are operated using CW light. (b) Pros and cons of the pulsed light and CW light for quantum state generation using heralding scheme and for measurement-based quantum processor. }
\label{fig:PulseCWarchitecture}
\end{figure}

Light is a promising platform for quantum computing thanks to the scalability achieved through time-domain multiplexing \cite{TDM_menicucci, TDM_menicucci_PRL}. 
In particular, in continuous-variable (CV) optical quantum computing, which uses the amplitude and phase of light as information carriers, quantum entanglement can be generated deterministically using linear optics and squeezed vacuum sources, and a large-scale quantum computing platform capable of performing multi-mode Gaussian operations has been experimentally realized \cite{Waritcluster, Mikkelcluster,MikkelmultimodeGauss,RIKENfullstack}.
It is known that non-Gaussian states are also required for universal quantum computing and error correction \cite{GKPoriginal,Wignernegative,Bartlett}, and recent reports on the generation of exotic non-Gaussian states using heralding methods \cite{Endo4photon, XanaduGKPpaper} have raised expectations for the realization of fault-tolerant optical quantum computing.
In this study, as shown in Fig.\ref{fig:PulseCWarchitecture}, we propose an architecture that integrates a non-Gaussian state generator with a cluster state and homodyne measurement system, which is a platform for large-scale universal quantum computation. In addition to that, we demonstrate a proof of principle of the hybrid architecture.

In previous studies of this field, either pulsed light \cite{Ourjoumtsev2006, Silberhornsinglephoton, Ra2020, Roh2025} or CW light \cite{Nehra:19, Darras2023, Konno2024, Yoshida2025, Waritcluster} has been used. Each light source has it's own advantages as shown in Fig.\ref{fig:PulseCWarchitecture} (b). Pulsed light can be used to generate non-Gaussian states in short optical wave packets by taking advantage of its temporal localization. In fact, non-Gaussian states have been reported to be generated on wave packets as short as picoseconds and femtoseconds \cite{Sonoyama:24,Cooper:13,Bouillard:19}. In time-domain multiplexing techniques, the wave packet width determines the upper limit of the quantum computing clock frequency, so achieving shorter wave packets is extremely important for realizing high-speed quantum computing. In addition, non-Gaussian states can be generated by a heralding method using photon-number-resolving detectors (PNRDs), and the use of pulsed light allows multiple photons to be detected at the same time, making it compatible with a PNRD. However, when pulsed light is used, it is difficult to achieve higher efficiency for the interference because of the temporal-mode mismatch. In cluster states generation and homodyne measurement, the interference of light by a beam splitter plays a major role. Thus, pulsed light is not compatible with quantum processors using cluster states and homodyne measurement. The opposite is true when CW light is used. Low-loss quantum entanglement generation and homodyne measurements are easily realized. On the other hand, in non-Gaussian state generation, the photon detector's timing jitter limits the optical wavepacket's temporal width to the order of sub-nanoseconds to nanoseconds \cite{SonoyamaPRR, Kawasaki2024}, limiting the clock frequency. Considering these characteristics, we propose a hybrid architecture that achieves both high speed and low loss by using pulsed light for non-Gaussian state generation and CW light for cluster state generation and homodyne measurement. 

As the first demonstration of the proposed pulse-CW hybrid quantum information processing, a homodyne measurement using a CW local oscillator for single-photon states generated by pulsed light is implemented. 
The estimated temporal width of the wavepacket of the single-photon state is as short as about 74 ps, which is approximately one to three orders of magnitude shorter than the previous researches using CW light resources \cite{Kawasaki2024, SonoyamaPRR}. 
The fidelity to the ideal single-photon state is around 74\%, and the value of the Wigner function at the origin is $W(0,0) = -0.153 \pm 0.003$. CV cluster states with non-Gaussian resources are known to be able to achieve fault-tolerant and universal quantum computation \cite{PRXQuantum.2.030325}. By utilizing the hybrid pulse-CW technique demonstrated in this experiment, we expect the realization of an ultrafast and low-loss fault-tolerant optical quantum computation.

\section*{Results}
The quadratures $x$ and $p$ of light used as carriers of information in CV optical quantum information processing are defined as $\hat{x}=(\hat{a}+\hat{a}^{\dagger})/\sqrt{2}$, $\hat{p}=(\hat{a}-\hat{a}^{\dagger})/\sqrt{2}i, (\hbar=1)$ using photon creation and annihilation operators $\hat{a}^{\dagger}$ and $\hat{a}$. 
$\hat{x}$ and $\hat{p}$ are Hermitian operators. 
They are conjugate physical quantities that are measurable but cannot be determined simultaneously because they satisfy an uncertainty relation. 
Experimentally, they can be measured by homodyne measurement, in which classical coherent light is interfered with signal quantum light by a beam splitter, and detected by a balanced detector. 
The optical quantum state in which we are interested is defined in the optical wave packet, and the quadrature in the wavepacket mode is $\hat{x}_{f}=\int dt f(t) x(t)$ for the envelope shape function $f(t)$ of the wave packet.

\begin{figure}[ht]
\centering
 \includegraphics[width=\textwidth]{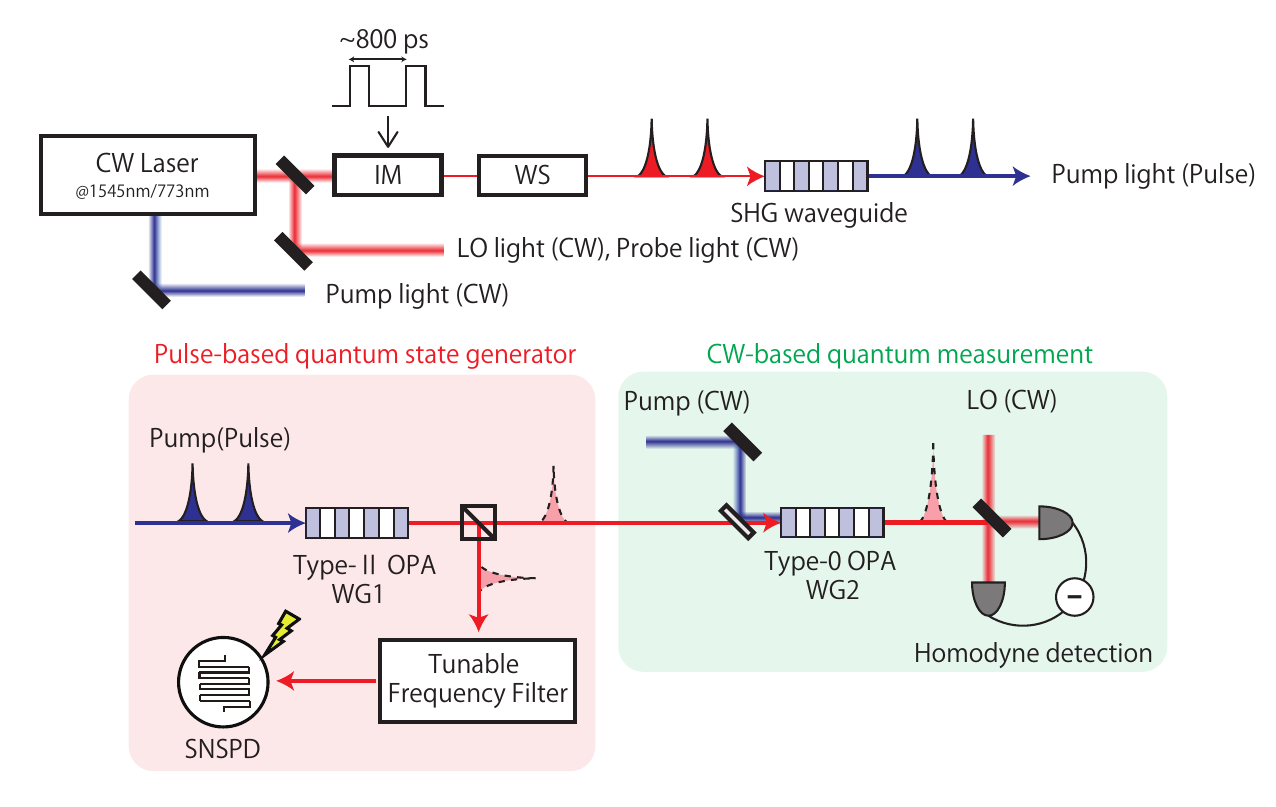}
 \caption{Schematic diagram of the experimental system. IM: Intensity Modulator, WS: Waveshaper, SHG: Second Harmonic Generation, LO: Local Oscillator, WG: Waveguide, OPA: Optical  Parametric Amplifier, SNSPD: Superconducting Nanostrip Photon Detector}
\label{fig:Experimentalsetup}
\end{figure}

A simplified diagram of the experimental system is shown in Fig.\ref{fig:Experimentalsetup}, where the pulsed pump light is generated with intensity modulation of CW light source. The pulsed pump light is injected into the Type-I\hspace{-1.2pt}I PPLN waveguide and the two-mode squeezed state is generated by parametric down conversion. When one of the generated two-mode squeezed states is detected by a photon detector, a single-photon state is generated in the other mode. For state verification, the quadrature of the generated quantum state is measured by homodyne measurement with CW local oscillator. 

The pulse width is several tens of picoseconds and has a bandwidth on the order of GHz. In this case, a homodyne measurement using a single-frequency CW light as a local oscillator requires a broadband homodyne measurement system that can directly measure GHz-bandwidth electrical signals. Conventional homodyne measurement systems have a bandwidth in the order of 100 MHz, but recently, a bandwidth of several tens of GHz has been achieved while maintaining high quantum efficiency by combining optical parametric amplification \cite{Inoue2023, Kawasaki2024, Kawasaki2025}. This technique of high-speed homodyne measurement is also used in this experiment. The quadrature of the quantum light to be measured is first amplified by optical parametric amplification, and then interfered with a local oscillator. It is then introduced into a low-efficiency but broadband balanced detector used in the field of optical communications. If the parametric gain is large enough, the loss after amplification is negligible, thus enabling measurement of high quantum efficiency regardless of the efficiency of the detector. In this experiment, the quadratures of the generated quantum states were obtained from 49455 homodyne measurement data, and the Wigner function was estimated by quantum state tomography \cite{Lvovsky2004}.

\begin{figure}[ht]
\centering
 \includegraphics[width=\textwidth]{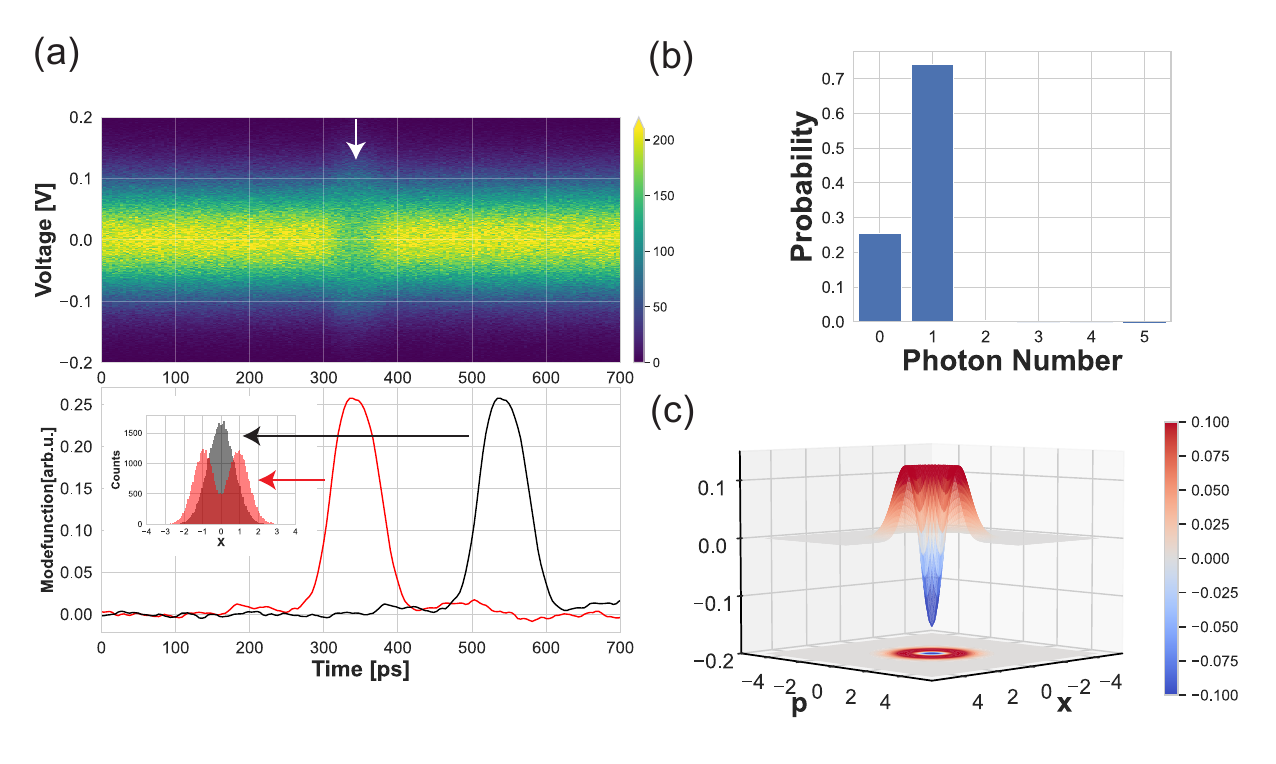}
 \caption{(a) The heatmap of homodyne measurement signals. The colorbar of the figure shows the relation between the color and the frequency of the 2D histogram. The section indicated by the white arrow corresponds to the generated single-photon state. The figure below shows the temporal mode function obtained from the homodyne measurement results by principal component analysis and its time shift. The quadratres corresponding to these temporal modes are calculated and plotted as a histogram in the figure on the left. (b,c) Photon number distribution and Wigner function estimated by quantum state tomography from the calculated quadrature data of the generated quantum state.}
\label{fig:results}
\end{figure}

The output signal of the 43 GHz balanced photodetector (Coherent, BPDV2150R-VF-FA) was read out by a broadband real-time oscilloscope (Keysight, UXR1104A), whose analog bandwidth is 110 GHz and sampling frequency is 256 GSa/s. The top figure of Fig.\ref{fig:results} (a) shows the heatmap of the homodyne measurement signals when photon detection occurred. Here, the voltage value at each time corresponds to the quadrature $\hat{x}(t)$ at each time, and the variance of homodyne signals increases at the time indicated by the white arrow. 
The red line in bottom figure of Fig.\ref{fig:results} (a) shows the temporal mode function $f(t)$ estimated by principal component analysis \cite{PCA_Lvovsky}. 
The full width at half maximum (FWHM) is about 74 ps.
From the estimated temporal mode function $f(t)$ and the value of the quadrature $\hat{x}(t)$ at each time, the quadrature $\hat{x}_f = \int dt f(t) \hat{x}(t)$ is calculated. 
A histogram of the quadrature $\hat{x}_f$ is shown in the left bottom in Fig.\ref{fig:results} (a). 
At the same time, the mode function with $f(t)$ shifted by approximately 200 ps on the time axis is shown as a black line and its quadrature histogram is plotted. It is checked that the variance of the quadratures in the temporal mode shifted by approximately 200 ps is identical to that of the quadratures in the temporal mode shifted by approximately 1 ns, which we use as a shot noise. Thus, we can say that the generated single-photon states are localized in several hundreds of picoseconds.

The density matrix and Wigner function of the generated states are estimated from the obtained quadrature data by the maximum likelihood estimation method \cite{Lvovsky2004}. Fig. \ref{fig:results} (b) shows the estimated photon number distribution of the generated state, The zero photon component accounted for 25.5\%, the single photon component for 74.0\%, and the two photons component for 0.5\%. This implies the pump for state generation was sufficiently weak and there were very few multi-photon detection events. The estimated Wigner function is shown in Fig.\ref{fig:results} (c), with the value $W(0,0) = -0.153 \pm 0.003$ at the origin. The fidelity to the single-photon state was $74\%$. To the best of our knowledge, the estimated Wigner function in this study has the most negative value at the origin and the highest fidelity to the single-photon state measured via homodyne tomography among the previous studies of pulsed light-based state generation. 
This may reflect the low-loss nature of this hybrid architecture. As for the optical wavepacket width, it is shorter by one to three orders of magnitude than those of previous studies generated with a CW light protocol \cite{Kawasaki2024, SonoyamaPRR}. 
From these results, we conclude that the strengths of the pulse-CW hybrid architecture, i.e., both high speed and low loss, have been demonstrated. 
Pulsed light is also compatible with PNRDs, and this hybrid architecture is highly scalable to exotic non-Gaussian states for which multi-photon detection is required. 
Recently, CW light-based broadband squeezed vacuum states and quantum entangled states have been demonstrated using waveguide OPAs \cite{Inoue2023,Kawasaki2025}. 
These are broader than the GHz bandwidth achieved in this study and can be combined to enable hybrid-type measurement-based quantum information processing. 
This architecture fully utilizes the advantages of both pulsed and CW light, and resolves the trade-off between high speed and low loss, which has been an issue in previous studies. 
We conclude that this is an important result for the realization of high-speed optical quantum information processing.

%変更予定(20250922追記)

\section*{Methods}
\subsection*{Phase Control}
Since the single-photon states generated in this study are phase insensitive, the phase between the pump light for state generation and the local oscillator (LO) light for homodyne measurement doesn't need to be locked. 
On the other hand, in the homodyne measurement using parametric amplification, the phase of the parametric amplification should be locked with the phase of the homodyne measurement. 
Therefore, in this experiment, a CW probe light for phase control was used in addition to the pulsed pump light for state generation and the CW LO light for homodyne measurement. 
The probe light passes through the same path as the quantum light and is injected into the homodyne measurement system. 
However, when the classical light is injected into the homodyne measurement system and photon detector during the measurement, the signal of the quantum light to be measured is buried. 
Therefore, in this study, we used a sample-and-hold method in which the measurement step (probe light off) alternates with the phase control step (probe light on). The bias current of the SNSPD was turned off during the phase control step to prevent the output from saturating due to the detection of classical light by the SNSPD. 

\subsection*{Details of the experimental system}
The laser source outputs CW light with a fundamental wavelength of around 1545 nm and a second harmonic of around 773 nm. 
CW light at 1545 nm is divided into three parts. One is the probe light for phase control, another is the LO light for homodyne measurement, and the last is the source of the pulsed light used for state generation. 
The pulsed light was generated by driving an intensity modulator (Exail, MXAN-LN-10) with a square wave whose repetition rate is around 1.25 GHz. 
The bandwidth was then limited by a wave shaper (Coherent, Waveshaper 1000A) and amplified by a two-stage Erbium-Doped-Fiber-Amplifier (EDFA) before being introduced into the SHG crystal (NTT Innovative Devices, WH-0773-000-A-B-C). The generated second-harmonic pulsed light was used as pump light for state generation, and the remaining fundamental pulsed light was used as a time reference for homodyne measurements. 
The second-harmonic CW light output from the laser was used as the pump light for parametric amplification in the first stage of the homodyne measurement.

The temporal mode of the generated quantum state depends not only on the waveform of the pump light and the phase matching function of the nonlinear optical crystal used to generate the state, but also on the wavelength filter of the idler. 
The state generation process using pulsed light is usually a multimode process in the frequency domain, and the photon detection mode and the state generation mode can be selected by narrowing the width of the idler's wavelength filter. 
In general, the idler and the signal are frequency correlated \cite{Silberhornsinglephoton}, and the center wavelength of the wavelength filter affects the center wavelength of the signal. 
In this experiment, a variable wavelength filter (Alnair, CVF-300CL) was used to set the center wavelength and the wavelength width. The center wavelength was set to around 1545.4 nm and the bandwidth was set to 0.03 nm.

\subsection*{Post-processing of measured data}
In this experiment, the signal from the photon detector was used as a trigger, and the output signal from the homodyne detector and the signal from the high-speed photodetector receiving pulsed light were acquired as time references. 
The measured raw signals have a timing jitter from frame to frame due to the finite temporal resolution of the SNSPD (about several tens of picoseconds). 
Therefore, the time was corrected by referring to the pulsed light signal received by the fast photodetector in each frame, which we belive have negligible timing jitter. 
In addition, the data processing omitted the frames in which the timing of pulsed light differs significantly from the timing of SNSPD detection in order to prevent the influence of dark counts. 
See the Supplement material for details.

\section*{Acknowledgement}
This work was partly supported by Japan Science and Technology (JST) Agency (Moonshot R\&D) Grant No. JPMJMS2064, the UTokyo Foundation, and donations from Nichia Corporation. T.Sonoyama acknoledges the funding from Japan Society for the Promotion of Science (JSPS) KAKENHI (No. 23KJ0518). K. Takase acknoledges the funding form JSPS (No. 23K13038). M. E. acknowledges the funding from JST (JPMJPR2254). T.Sonoyama, T.Suzuki, K.Takahashi, T.N., A.K. acknowledge financial supports from The Forefront Physics and Mathematics Program to Drive Transformation (FoPM), a World-leading Innovative Graduate Study (WINGS) Program, the University of Tokyo. A.K. acknowledges financial support from Leadership Development Program for Ph.D (LDPP), the University of Tokyo. K.Takase, W.A. and M.E. acknowledge supports from the Research Foundation for Opto-Science and Technology. This paper partially utilized AI translation and generative AI, such as Google Gemini (2.5 Flash), for the purposes of English translation, proofreading, and text organization. We also thank Mr. Katsuki Nakashima from the University of Tokyo for the fruitful discussion, Mr. Quentin Civario from OptQC Corp for providing assistance with English grammar and proofreading. 

\section*{Author contributions}
T.Sonoyama led the experiment with supervision from K.Takase., W.A., M.E., and A.F.. T.Sonoyama and K.Takahashi constructed the experimental setup for the single-photon state generation. T.Sonoyama, T.Sano, T.Suzuki constructed the experimental setup of the broadband homodyne systems. T.Sonoyama and T.Sano analyzed the acquired data. The discussions regarding the experiment were mainly conducted by T.Sonoyama, T.Sano, T.Suzuki, K.Takahashi, T.N, A.K, W.A, K.Takase, and M.E.. T.K., A.I., and T.U. provide the OPA used in the experiment. M.Y., S.M., and H.T. provide the SNSPD used in this experiment. T.Sonoyama wrote the manuscript with assistance from all the coauthors.

\section*{Competing interests}
Authors declare no competing interests.

\section*{Data and materials availability}
All data are available either in the manuscript or in the supplementary material.

\clearpage

% --- Supplementのカウンタ設定（図をS1, S2などにする場合） ---
\setcounter{section}{0}
\setcounter{figure}{0}
\setcounter{table}{0}
\setcounter{equation}{0}
\renewcommand{\thefigure}{S\arabic{figure}}
\renewcommand{\thetable}{S\arabic{table}}
\renewcommand{\theequation}{S\arabic{equation}}
\renewcommand{\thesection}{S\arabic{section}}

% --- 手動タイトル部分 ---
\begin{center}
    \textbf{\Large Supplementary Material for: \\ [0.5em] 
    Hybridization of pulse and continuous-wave \\ based optical quantum computation}
\end{center}

\vspace{2em}
\section*{Loss Budget}

\begin{table}[h]
\begin{center}
\begin{tabular}{c|c}
   & Effective loss(\%) \\ \hline
  Internal loss in the WG-1 & 6.7\% \\ 
  Propagation loss & 2.1\% \\ 
  Internal loss in the WG-2 & 1.0\% \\
  Loss after the WG-2 & 0.6\% \\ 
  Circuit noise of HD & 0.8\% \\
  Dark and Fake counts & 0.4\% \\
  Others & 16.1\% \\ \hline
  In total & 25.5\%
\end{tabular}
\caption{Loss Budget}
\label{tb:Lossbudget}
\end{center}
\end{table}

The single photon states generated are subject to loss due to various factors. Table.\ref{tb:Lossbudget} summarizes them. First of all, we call the waveguide OPA for state generation as waveguide-1 (WG-1) and the waveguide OPA for parametric amplification before homodyne detection as waveguide-2 (WG-2).
The internal loss of 6.7 \%  (WG-1) for state generation is estimated from a typical propagation loss value of 0.1 dB/cm for a PPLN waveguide \cite{Kashiwazaki2021}.
Next, the propagation loss of 2.1 \% is estimated from the propagation efficiency of classical light (probe light), which is in the same wavelength band as the generated state. 
Note that, this is the propagation loss after the PBS, which separates the signal light from the idler light, to just before the WG-2. Thus, this does not include the loss from the lens immediately after the WG-1.

The effective internal loss of 1.0\% in the WG-2 was then estimated from the propagation loss of the waveguide 7\% \cite{Kashiwazaki2021} and parametric gain of around 30 dB.
The mathematical discussion is given in the next section, but the effective loss to the quadrature of the quantum state is reduced, since the generated state is subjected to loss while being amplified. 
The effective loss after parametric amplification was estimated to be reduced from approximately 90 \% to 0.6\% thanks to the parametric gain of about 30 dB. 90\% loss include 26.1\% propagation loss, 40.1\% loss due to fiber coupling, 11.5\% loss due to mode matching with LO light in the homodyne measurement system, and approximately 70\% loss due to the photodiode's efficiency of the homodyne detector.
A mathematical discussion is given in the next section.
The effective loss 0.8\% due to circuit noise was estimated from the shotnoise and electrical circuit noise clearance. 
Finally, the 0.4\% loss due to dark counts and fake counts in photon detection was estimated by comparing the counts when the pump light for state generation was input with the counts when it was not input. 
Note that this table does not cover all sources of loss. 
The largest loss we are not considering could be the coupling loss from free space to the WG-2 for parametric amplification. The loss value estimated from the probe light coupling to WG-2 was approximately 26\%. However, this is not consistent with the single-photon state's fidelity we measured. Since the spatial modes of the probe light and quantum light do not necessarily match, this value is not exactly the effective loss of the generated quantum states. Also, measurements performed on a different day yielded a lower loss, thus we think the coupling loss estimated using probe light is not so reliable. Other than that, pulsed light state generation is a multimode process in the frequency domain, and its contribution to the effective loss is difficult to estimate experimentally. These are summarized in the Others section. \\
Finally, the total loss 25.5\% is from the proportion of zero-photon component which is experimentally estimated by homodyne tomography. From this value, the effective loss in the Others section are estimated.

\begin{figure}[ht]
\centering
 \includegraphics[width=\textwidth]{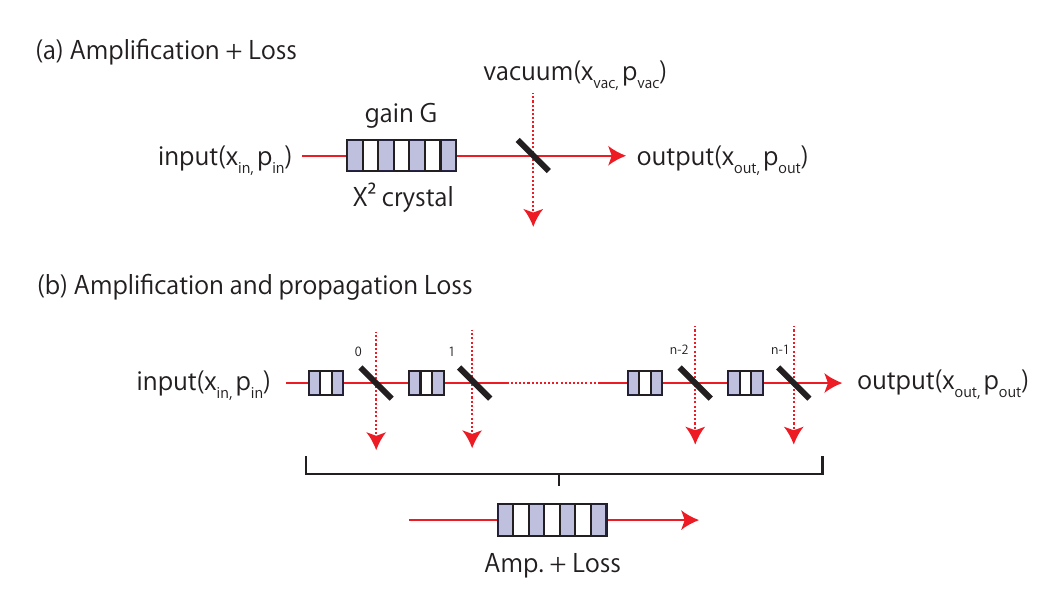}
 \caption{(a) The process of suffering loss after undergoing parametric amplification. (b) The process of suffering loss while undergoing parametric amplification.}
\label{fig:Parametricamploss}
\end{figure}

\section*{Treatment of optical loss during and after parametric amplification}
Fig.\ref{fig:Parametricamploss} (a) schematically illustrates the process of loss after parametric amplification. 
If the parametric gain is $G$ and the loss is $L$, the process of parametric amplification for the quadrature $\hat{x}$ can be expressed as follows.
% この後の式は20250812に追加
\begin{eqnarray}
  \hat{x}_{\rm out} &=& \sqrt{1-L}(\sqrt{G}\hat{x}_{\rm in}) + \sqrt{L}\hat{x}_{\rm vac} \\
  &=& \sqrt{(1-L)G+L}\left(\sqrt{1-L'}\hat{x}_{\rm in} + \sqrt{L'}\hat{x}_{\rm vac}\right)
\end{eqnarray}
From this formula, the effective loss $L'$ can be calculated as follows.
%,を.に変更(20260421). 以下equationをeqnarrayに変更(20260421)。
%この後の式も L' = のイコールの後を20250812に追加
\begin{eqnarray}
L’ = \frac{L}{(1-L)G+L}
\end{eqnarray}
As you can see, the effective loss $L'$ is reduced from the loss $L$ thanks to the parametric gain $G$.
Next, let us consider the process of loss during parametric amplification as shown in Fig.\ref{fig:Parametricamploss} (b). 
The input-output relation is expressed by the following equation. 
% この後の式は20250812に追加
% x_inが抜けていたのでそこを20250817に追加
\begin{eqnarray}
  \hat{x}_{\rm out} &=& \sqrt{(\eta g)^n} \hat{x}_{\rm in}+ \sum_{i=0}^{n-1} \sqrt{(\eta g)^{n-1-i}} \sqrt{1-\eta} \hat{x}_{i, {\rm vac}} \\
  &=& \sqrt{(\eta g)^n} \hat{x}_{\rm in}+ \sqrt{\frac{(\eta g)^n - 1}{\eta g -1} (1-\eta)} \hat{x}_{\rm vac}
\end{eqnarray}
Here, we are using the fact that the linear combination of independent vacuum states equal to another vacuum state. $\eta$ is the transmittance efficiecy and $g$ is the parametric gain of the segmented waveguide. Thus, $\eta^n = 1-L$, $g^n = G$ are satisfied. 
When we consider that the number of segments $n$ become large, we can derive the expression of the output quadrature as follows. 
% x_vacが追加されていない気がするので、20250817に追加。
\begin{eqnarray}
  \hat{x}_{\rm out} = \sqrt{G(1-L)}\hat{x}_{\rm in} + \sqrt{\frac{(1-G(1-L))\ln(1-L)}{\ln(G(1-L))}} \hat{x}_{\rm vac}
\end{eqnarray}
From this equation, the effective loss $L'$ is expressed as follows.
% この後の式も20250812に追加
% (eqnarray)となっていたところを{eqnarray}に直した。(20250817)
\begin{eqnarray}
  L' = \frac{(1-(1-L)G)\ln(1-L)}{(1-L)G\ln G + \ln(1-L)}
\end{eqnarray}
Using these equations, the effective loss in Table.\ref{tb:Lossbudget} was calculated.

% Captionが逆な気がするので、その部分は20250731に変更。

%変更予定
\begin{figure}[ht]
\centering
 \includegraphics[width=\textwidth]{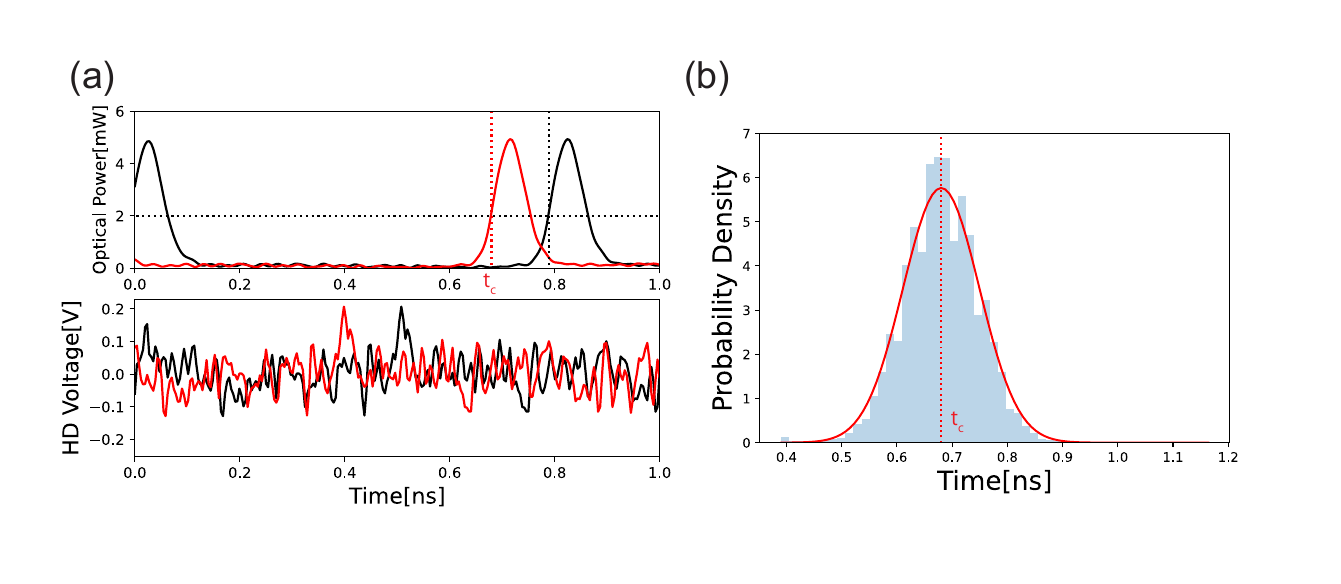}
 \caption{(a) Displays one frame of homodyne measurement signals and pulsed light signals received by a photodetector.
The black line is the raw signal, and the red line is the time-corrected signal. 
Time correction was performed so that the rise time of the pulsed light exceeded the threshold became $t=t_c$.
(b) Histogram of the rise times of the pulsed light in the raw signal. In the raw signal, the rise of the SNSPD is used as the time reference. The red line is a Gaussian fit of this. The center time of this is $t=t_c$. 
}
\label{fig:CorrectionofTD}
\end{figure}

%変更予定
\begin{figure}[ht]
\centering
 \includegraphics[width=\textwidth]{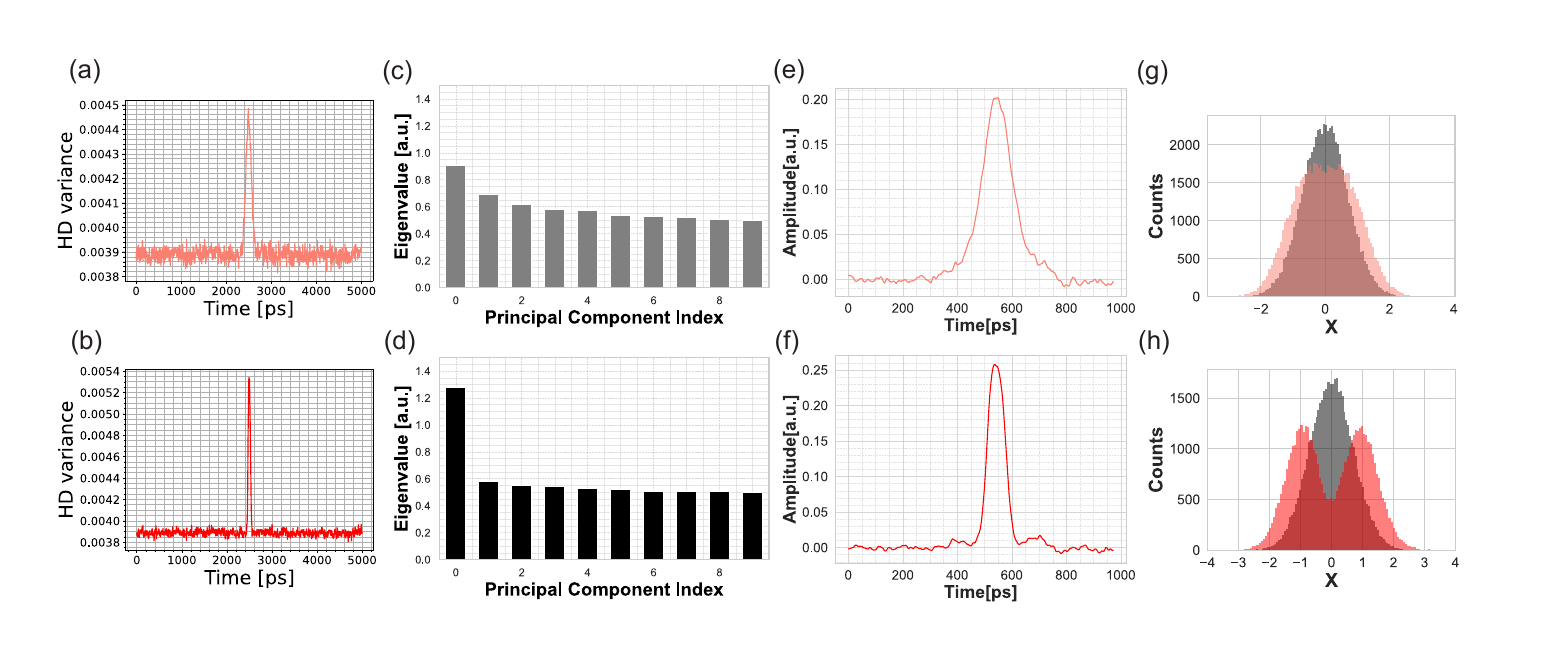}
 \caption{The results are based on an analysis using homodyne measurement data without the time correction in the upper row and with the time correction in the lower row.
From left to right: the variance of the homodyne measurement data at each time point, the principal components obtained by principal component analysis arranged in descending order, the first time mode function corresponding to the maximum principal component obtained by principal component analysis, 
and the histogram of the quadratures in the first time mode.}
\label{fig:SNSPDtrigcompare}
\end{figure}

\section*{Data post-processing}
In this measurement, the signal from the photon detector SNSPD was used as a trigger to acquire the output signal of the homodyne detector. And the signal of the pulsed light received by the high-speed photodetector is recorded with the homodyne signal to be used for the time reference. 
The timing jitter of the photon detector SNSPD is several tens of ps, which causes the signal of the homodyne measurement to be slightly off at each measurement. 
Considering that the wavepacket width of the single-photon state generated in this study is about 70 ps, the timing jitter of the SNSPD is not negligible. 
Therefore, we corrected the time later using the pulsed light signal received by a high-speed photodetector with a higher time resolution.
Fig.\ref{fig:CorrectionofTD} (b) shows a histogram of the rise time of the pulse signal with respect to the rise time of the SNSPD. 
The red line in Fig.\ref{fig:CorrectionofTD} (b) is the histogram fitted with a Gaussian function, and the center time is $t=t_c$. 
In the time correction, the deviation of the rise time of the pulse signal from $t=t_c$ is corrected as shown in Fig.\ref{fig:CorrectionofTD} (a). 
Fig.\ref{fig:SNSPDtrigcompare} shows the difference between the results of data analysis with and without the time delay correction. 
The upper figures correspond to the results without the time delay correction and the bottom figures correspond to the results with the time delay correction. 
As you can see, the generated states are scattered to multi temporal modes because of the timing jitter. 

\begin{figure}[ht]
\centering
 \includegraphics[width=12cm]{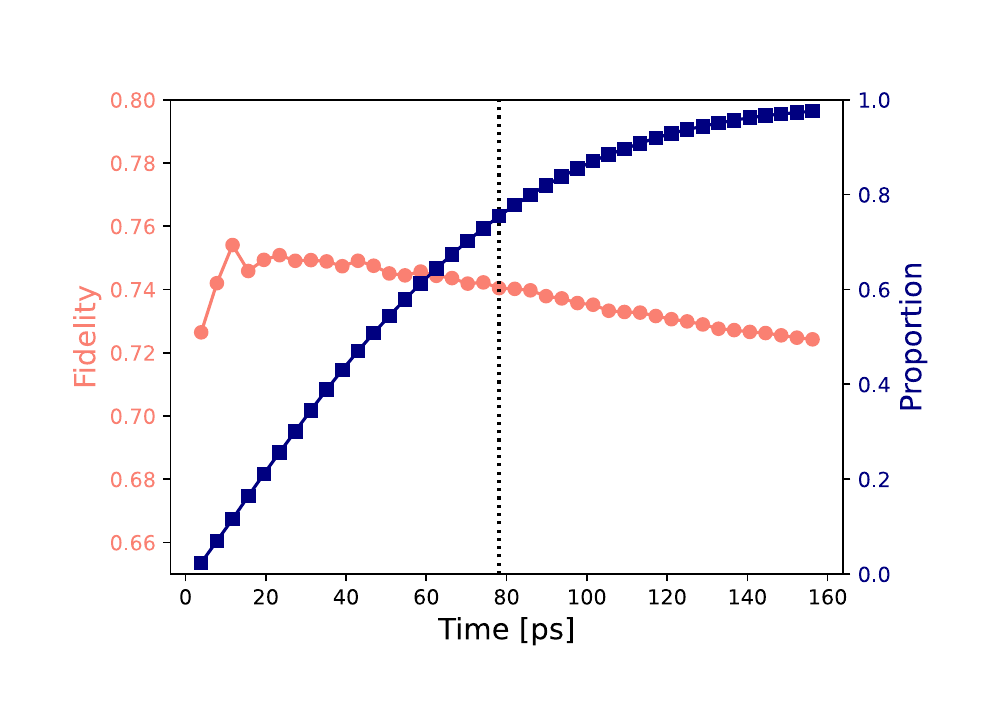}
 \caption{The ratio of data extracted during post-processing only when the rise time of the pulse light is within the range of $t_c \pm \Delta t$, and the fidelity with the single-photon state when performing quantum state tomography. The horizontal axis is $\Delta t$.}
\label{fig:FidelityandProportion}
\end{figure}

In addition to that, the data with a large deviation from $t=t_c$ in the rise time of the pulse signal were removed in the post-processing. 
There are two main reasons for this: first, to reduce the influence of dark counts; second, using events in which the rise time of pulsed light is close enough to $t=t_c$ may reduce multimodality in the time domain of the generated state. 
Here, the data is extracted when the rise time of the pulse signal is within the range of $t_c-\Delta t$ to $t_c + \Delta t$. 
As you can see in Fig.\ref{fig:FidelityandProportion}, when $\Delta t$ is decreased, the fidelity with the single-photon state improves. 
In the data analysis in the text, $\Delta t$ was set to about 80 ps (corresponding to the dotted line in Fig.\ref{fig:FidelityandProportion}), and the number of data points used in the analysis was about 75\% of the total number of data points.

%変更予定

\begingroup
\justifying % 柔軟な行揃えを適用（\RaggedRightでも可）
\sloppy
\clearpage
\printbibliography
\endgroup

\end{document}